\documentclass[12pt]{iopart}

\usepackage{graphicx}
\usepackage[sort&compress]{natbib}
\usepackage{dcolumn}
\usepackage{subdepth}
\usepackage{color}
\usepackage{enumitem}
\usepackage{blindtext}
\usepackage{soul}
\usepackage{hyperref}
\setstcolor{red}

\newcommand{\VCL}{V_{\mathrm{CL}}}

\newcommand{\VMBO}{V_{\mathrm{MBO}}}
\newcommand{\VS}{V_{\mathrm{S}}}
\newcommand{\VSt}{\tilde{V}_{\mathrm{S}}}

\newcommand{\Eq}[1]{Eq.~(\ref{#1})}
\newcommand{\Eqs}[1]{Eqs.~(#1)}
\newcommand{\Fig}[1]{Fig.~\ref{#1}}
\newcommand{\Sec}[1]{Sec.~\ref{#1}}

\newcommand{\cm}{cm$^{-1}$}
\newcommand{\diff}{\mathrm{d}}
\renewcommand{\i}{\mathrm{i}}

\bibpunct{[}{]}{,}{n}{}{;}

\begin{document}

\title{Vibrational spectroscopy via the Caldeira-Leggett model with anharmonic system potentials}

\author{Fabian Gottwald, Sergei D Ivanov$^*$, Oliver K\"uhn}
\ead{sergei.ivanov@uni-rostock.de}
\address{Institute of Physics, Rostock University, Albert Einstein Stra{\ss}e 23-24, 18059 Rostock, Germany}

\begin{abstract}
The Caldeira-Leggett (CL) model, which describes a system bi-linearly coupled to a harmonic bath, has enjoyed popularity in condensed phase spectroscopy owing to its utmost simplicity.
However, the applicability of the model to cases with anharmonic system potentials, as it is required for the description of realistic systems in solution,
is questionable due to the presence of the invertibility problem 
[\href{http://pubs.acs.org/doi/abs/10.1021/acs.jpclett.5b00718}{J.\ Phys.\ Chem.\ Lett., \textbf{6}, 2722 (2015)}]
unless the system itself resembles the CL model form.
This might well be the case at surfaces or in the solid regime, which we here confirm for a particular example of an iodine molecule in the atomic argon environment under high pressure.
For this purpose we extend the recently proposed Fourier method for parameterizing linear generalized Langevin dynamics
[\href{http://scitation.aip.org/content/aip/journal/jcp/142/24/10.1063/1.4922941}{J.\ Chem.\ Phys., \textbf{142}, 244110 (2015)}] to the non-linear case based on the CL model and perform an extensive error analysis.
In order to judge on the applicability of this model in advance, we give handy empirical criteria and discuss the effect of the potential renormalization term.
The obtained results provide evidence that the CL model can be used for describing a potentially broad class of systems.
\end{abstract}

\date{\today}
\submitto{\NJP}
\maketitle

\section{Introduction}
The analysis of complex dynamical processes in many-particle systems is
one of the main goals in modern physics and requires sophisticated experimental setups and reliable theoretical models.
Although to date computer facilities allow one
to treat an increasingly large amount of coupled degrees of freedom (DOFs),
a reduction of the description to few variables is convenient in many cases, 
since this can not only ease the interpretation, but enable the identification of key properties~\cite{Kuehn2011}.
Such a reduced description can formally be obtained from a so-called system-bath partitioning, where only a small subset of DOFs, referred to as system, is considered as important for describing a physical process under study.
All other DOFs, referred to as bath, are regarded as irrelevant in the sense that they might influence the time evolution of the system but do not 
\textit{explicitly} enter any dynamical variable of interest. 
Conveniently, reduced equations of motion (EOMs) for the system DOFs are derived, in which the influence of the bath
is limited to dissipation and fluctuations only.

To obtain a reduced description, different strategies can be employed. 
A very popular approach is to assume a simple form of the bath such that its DOFs can be easily integrated out from the system's EOMs.
Particularly, in the Caldeira-Leggett (CL) model, the environment is assumed to be a collection of independent harmonic oscillators, bi-linearly coupled to the system~\cite{Caldeira-PRL-1981,Caldeira-AP-1983,Zwanzig1973,ZwanzigBook2001}.
This model has been widely used in analyzing and interpreting (non-)linear spectroscopic experiments on systems in condensed phase, termed multi-mode Brownian oscillator (MBO) model in this context~\cite{Mukamel1995,Palese1996,Okumura1997,Woutersen1999,Toutounji2002,Tanimura2009,Joutsuka2011}.
The resulting reduced EOM is known as the generalized Langevin equation (GLE),
where the bath's influence is described by a frequency-dependent friction and a stochastic force with a finite correlation time.
Depending on the type of the CL model, the system potential is (or is not)
modified by a harmonic renormalization term, see \Sec{sec:CL model} for a discussion.

Another approach to reduced EOMs is to employ formal projection operator techniques to project out the bath from the system's EOM resulting in linear or non-linear GLE forms~\cite{Mori1965,Kawasaki1973,Kawai2011,ZwanzigBook2001}. 
In the former, the resulting system potential is effectively harmonic, whereas in the latter the system potential is formed by a (non-linear) mean-field potential.
In both cases, noise and dissipation can be mathematically defined as (non-)linearly projected quantities.
Independently on the derivation, the general advantage of the GLE
is that the dissipation and the statistical properties of the noise are entirely described by the so-called memory kernel being simply a function of time or, equivalently, by the spectral density as its frequency domain counterpart.
Due to this simplicity, the GLE has been applied in many fields as, for instance, in the theory of vibrational relaxation for estimating characteristic relaxation times~\cite{Whitnell1990,Benjamin1993,Tuckerman1993a,Gnanakaran1996a},
reaction rates~\cite{Grote-JCP-1980} and for thermostatting purposes~\cite{Ceriotti2009,Ceriotti2009a, Ceriotti2010};
see, e.g., Refs.~\cite{Abe-PR-1996,Tanimura2006,Snook-Book} for review.

Despite its popularity, establishing a connection between the CL model (or the respective GLE) and a real many-particle system is not straightforward.
In Ref.~\cite{Gottwald-JPCL-2015} we have shown that it is generally not possible to establish an invertible mapping between the two due to the so-called invertibility problem.
In a nutshell,
the memory kernel is unambiguously determined by momentum--momentum and momentum--system force correlation functions that are available,
e.g.\ directly from molecular dynamics (MD) simulations.
It turned out that there exist infinitely many pairs of such correlation functions that correspond to the same memory kernel
and, hence, the corresponding GLE cannot mimic the true system dynamics.
There are two important exceptions, however: i) the CL model's system potential is taken effectively harmonic; ii) the real system is (approximately) of the CL form, that is the bath is (almost) harmonic and the coupling to the system is (almost) bi-linear.
The former is not satisfactory if we are interested in non-linear spectroscopy, though can be well applied to study dissipation dynamics as well as linear response properties.
Also the price to pay is that any connection to atomistic description is sacrificed, since all the anharmonicity in the 
system is projected onto the bath.

The second exception, however, suggests that there might exist a broad class of systems which satisfy these criteria.
Unfortunately, we have shown that this is not the case for typical molecular systems in solution, such as
aqueous systems and the ionic liquid [C$_2$mim][NTf$_2$]~\cite{Gottwald-JPCL-2015}.
Still, considering the somewhat  artificial A$_2$ in A system adopted from Ref.~\cite{Berne1990} showed that it can be
satisfactorily mapped onto the model~\cite{Gottwald-JCP-2015}.

Importantly, if such a mapping can be established, then the full quantum-mechanical treatment of the bath can be performed analytically via the Feynman-Vernon influence functional~\cite{Feynman-AP-1963,Caldeira-PhysicaA-1983} without any further approximations.
Moreover, numerically exact hierarchy type equation of motion approaches
are essentially based on the CL model~\cite{Dijkstra-PRL-2010,Suess:2014gz}.
Further, it is ideally suited for numerical methods that solve the Schr\"odinger equation in many dimensions~\cite{giese06_211}.
The CL model can also be taken as a starting point for quantum-classical hybrid simulations, that is by treating only the usually
low-dimensional system part quantum-mechanically and the bath (semi)classically~\cite{egorov99_5238}.
Finally, the machinery for a purely classical treatment by means of a GLE is provided by the method of colored noise thermostats~\cite{Ceriotti2009,Ceriotti2009a,Ceriotti2010}.
This means that if the mapping were established then the CL model would provide a unified framework for reduced dynamics as well as for the quantum-classical comparison of dynamical properties of real molecular systems.

In this paper, we aim at finding cases when the anharmonic CL model,
that is the CL model with the anharmonic system potential resulting in the corresponding non-linear GLE,
is successful in describing a many-particle system and formulating the criteria when one can expect such a success.
Further, we extend our recently suggested Fourier-based scheme for calculating memory kernels from explicit MD
simulations for the harmonic cases~\cite{Gottwald-JCP-2015} to the anharmonic ones.
Since the resulting protocol is much more sensitive to the numerical accuracy of the input data than the one for the linear GLE, a comprehensive error analysis is provided as well.

The paper is organized as follows.
In \Sec{sec:CL model} we review the basic theory underlying the CL model.
Section \ref{sec:method} contains the detailed description of the proposed numerical scheme for computing the spectral density including the error analysis.
After giving the computational details in \Sec{sec:comp_det} we 
apply the developed scheme to a specific system where the mapping onto the CL model with an anharmonic system potential is satisfactory in \Sec{sec:applications}.
Further we discuss the possibility to formulate the criteria for an existence of a mapping between the CL model and the real molecular system followed by a general conclusion in \Sec{sec:conclusion}.

\section{The Caldeira-Leggett Model\label{sec:CL model}}

The CL model has enjoyed great popularity in condensed phase spectroscopy~\cite{Palese1996,Okumura1997,Woutersen1999,Toutounji2002,Tanimura2009,Kleinekathoefer-JPCD-2010,Joutsuka2011}.
It assumes that the bath consists of independent harmonic oscillators
bi-linearly coupled to the system.
The total potential energy for the model reads
\begin{equation}
\label{eq:CL Hamiltonian}
\VCL(x,\{Q_i\})=\VS(x)+\sum_{i}\frac{1}{2}\omega_{i}^{2}Q^2_{i}-\sum_{i}g_{i} Q_i x 
\end{equation}
with the bath frequencies $\omega_{i}$, bath masses set to unity and the coupling strengths
$g_{i}$~\cite{Caldeira-PRL-1981,Caldeira-AP-1983,grabert88_115}.
For the sake of presentation we assume the system to be one-dimensional.
One might complete the square for the bath coordinates, resulting in
\begin{equation}
\label{eq:CL Hamiltonian renormalized}
\VCL(x,\{Q_i\})=\VSt(x)+\sum_{i}\frac{1}{2}\omega_{i}^{2}\left (Q_{i} - \frac{g_i}{\omega_i^2} x\right )^2
\end{equation}
with the renormalized system potential
\begin{equation}
\label{eq: corrected potential}
\VSt(x) \equiv \VS(x) - \frac{1}{2} \sum_i \frac{g_i^2}{ \omega_i^2} x^2
\enspace ;
\end{equation}
note that this renormalization emerges in the EOMs irrespectively whether \Eq{eq:CL Hamiltonian} or \Eq{eq:CL Hamiltonian renormalized} is used as a starting point~\cite{Cortes1985a}.
The two potentials, \Eqs{\ref{eq:CL Hamiltonian},\ref{eq:CL Hamiltonian renormalized}}, are mathematically equal, and differ exclusively by the way of partitioning into the system and the bath,
which may become important if the two are treated differently.

%
The physical consequences of such a renormalization of the system potential have been addressed at different places in the literature. 
%
Petruccione and Vacchini considered the necessity of the counter term from the standpoint of obtaining a translationally invariant reduced dynamics for the Brownian particle in a homogeneous fluid~\cite{Vacchini-PRE-2005}.
Wisdom and Clark have discussed the renormalization effect in the context of enhancing tunneling probabilities through a barrier~\cite{Widom-1982-63,Widom-1982-1572}.
%
The quasi-adiabatic propagator for path integral methods (QUAPI) developed by Makri and coworkers profited from the presence of the renormalization term for the description of reaction dynamics~\cite{Makri-CPL-1993, Makri-JCP-1994}.
Caldeira and Leggett have given insight into when and when not to expect a renormalization effect~\cite{Caldeira-AP-1983}.
According to them, one can expect it in nuclear dynamics, for instance, when a collective DOF is coupled to many single particle modes.
In contrast, no renormalization should be expected if the system-bath coupling is adiabatic. 
Additionally, the authors have shown that a rigorous treatment of systems in an electromagnetic field can be described by a CL Hamiltonian with a counter term that cancels the renormalization of the potential~\cite{Caldeira-AP-1983}.
Removing the renormalization term from the potential amounts to
\begin{equation}
\label{eq:MBO Hamiltonian}
\VMBO(x,\{Q_i\})=\VS(x)+\sum_{i}\frac{1}{2}\omega_{i}^{2}\left (Q_{i} - \frac{g_i}{\omega_i^2} x\right )^2
\enspace .
\end{equation}
This form of the CL model is often termed multi-mode Brownian oscillator (MBO) model~\cite{Mukamel1995}.
Here we will adopt this terminology and refer to \Eq{eq:CL Hamiltonian renormalized} and to \Eq{eq:MBO Hamiltonian} as to the CL and the MBO model, respectively.
Note that there are no differences between the models, apart from the presence/absence of the
aforementioned renormalization term.

Given the similarity of the models, a GLE can be derived from either of them both in the classical~\cite{Zwanzig1973,Cortes1985a,ZwanzigBook2001} and 
quantum~\cite{Ford-JSP-1987,Ford-PRA-1988} domains; the respective subscripts (CL or MBO) will be omitted from now on unless there would be a need for explicit distinction.
Limiting ourselves to the classical description, the derivation can be straightforwardly performed
without any further approximations by integrating the EOMs for the bath, yielding
\begin{equation}
\label{eq:CL-GLE}
\dot{p}(t) = 
F[x(t)]-
\intop_{0}^{t}\xi(t-\tau)p(\tau)\diff \tau+R(t)
\enspace ,
\end{equation}
where the system force $F[x(t)]\equiv-\partial_x \VSt[x(t)]$ for the CL model or
$F[x(t)]\equiv-\partial_x \VS[x(t)]$ for the MBO model.
In the GLE the bath is reduced to non-Markovian dissipation and fluctuations,
represented by the memory kernel, $\xi(t)$, and the so-called noise term or fluctuating term, $R(t)$, respectively.
The memory kernel can be written down explicitly as 
\begin{equation}
\xi(t)  = \sum_{i}\frac{g_{i}^{2}}{m\omega_{i}^{2}}\cos(\omega_{i}t)
\enspace ,
\end{equation}
where $m$ is the mass of the system DOF.
Although it is possible to write down a cumbersome explicit expression for $R(t)$,
it is usually employed via a stochastic model~\cite{ZwanzigBook2001}.
The form of this explicit expression dictates that the fluctuating force is gaussian-distributed and has zero mean.
Its time correlations are described by means of the fluctuation-dissipation theorem (FDT) that connects the fluctuating force with the memory kernel, that is with dissipation
\begin{equation}
\label{eq:FDT}
\left\langle R(0)R(t)\right\rangle =mkT\xi(t)
\enspace ,
\end{equation}
thereby establishing the canonical ensemble with the temperature $T$.
Defining the spectral density as the one-sided Fourier transform (denoted with a hat)
from the memory kernel yields
\begin{equation}
\label{eq:MemoryKernel}
\Re \hat{\xi}(\omega) \equiv \Re \int\limits_0^\infty \diff t \, \e^{-\i \omega t}\xi(t)=
\sum\limits_i^\infty \frac{g_i^2}{m \omega_i^2} \delta(\omega - \omega_i)
\enspace .
\end{equation}
Comparing this to \Eq{eq: corrected potential} immediately gives
\begin{equation}
\label{eq:renormalized Potential}
\VSt(x)=\VS(x) - \frac{1}{2}m x^2 \int\limits_0^\infty \Re \hat \xi (\omega)\diff \omega 
\enspace .
\end{equation}
It is worth mentioning again that the definitions of the spectral density, the memory kernel and the FDT, coincide for the GLEs resulting from the CL and MBO models.
Nevertheless, the system dynamics is different and the choice of the model is system-dependent.
Therefore, we employ both models and compare their performance against each other in the following.

\section{Calculating the spectral density \label{sec:method}}

\subsection{Fourier method}
\label{sec:Fourier}
Recently, we argued that the problem of parameterizing the memory kernel naturally poses itself in the 
frequency domain and proposed a method for parameterizing linear (Mori-Zwanzig) GLEs, i.e.~for harmonic system potentials~\cite{Gottwald-JCP-2015};
%
note that a similar in spirit but technically different method for linear GLEs was developed independently~\cite{Bonfanti-AdP-2015}
and applied to quantum dynamics of hydrogen atoms on graphene~\cite{Bonfanti-JCP-2015,Bonfanti-JCP-2015-2}.
In the following, we extend the method to the non-linear GLEs, \Eq{eq:CL-GLE}, based on the CL (or MBO) model with an anharmonic system potential.
The starting point is the integro-differential equation for the momentum autocorrelation function (MAF), $C_{pp}(t)\equiv \langle p(t)p(0) \rangle$ and the correlation function of momentum and system force (MFC), $C_{pF}(t)\equiv \langle F(t)p(0) \rangle$,
\begin{equation}
\label{eq:Volterraequation-CL}
\dot{C}_{pp}(t)= C_{pF}(t) - \intop_0^t\xi(t-\tau)C_{pp}(\tau) \diff \tau
\enspace ,
\end{equation} 
where the system force, $F(t)$, is defined with respect to the underlying model, see above.
This equation can be derived by multiplying the GLE, \Eq{eq:CL-GLE}, with $p(0)$ followed by a canonical average over initial conditions; note that the initial momenta are uncorrelated with the noise.
To obtain the memory kernel one can invert \Eq{eq:Volterraequation-CL} taking the MAF and the MFC, computed from explicit MD simulations, as an input.
Following the derivation of the Fourier method for linear GLEs, we perform
a half-sided Fourier transform of \Eq{eq:Volterraequation-CL} resulting in
\begin{equation}
i\omega \hat{C}_{pp}(\omega)-C_{pp}(t=0)=\hat{C}_{pF}(\omega)-\hat{\xi}(\omega)\hat{C}_{pp}(\omega)
\enspace .
\end{equation}
Note that the term $C_{pp}(t=0)$ stems from the half-sided transform and would vanish if a full-sided Fourier transform were taken.
This algebraic equation can be directly solved for the spectral density,
\begin{equation}
\label{eq:ComputingKernel}
\hat{\xi}(\omega) = \frac{1+\hat{C}_{pF}(\omega)}{\hat{C}_{pp}(\omega)} -i\omega
\enspace ,
\end{equation}
where it has been assumed that both, the MAF and the MFC, are normalized to $C_{pp}(t=0)$.
Calculating the spectral density for the MBO model according to \Eq{eq:ComputingKernel} is straightforward as all the required ingredients, the MAF and the MFC, are available from MD simulations directly.
In contrast, for the CL model one has to compute an MFC with a renormalized system force, where the renormalization is not known a priori.
Nonetheless, since the spectral density is exactly the same in both models, one can always compute it for the MBO model and use it within both models.
The renormalization term can be then computed from the obtained spectral density according to \Eq{eq: corrected potential}.
Note that the Fourier method for linear GLEs 
requires just 
the MAF from which the true kernel, $\xi(t)$, and the effective system frequency can be fitted, see Eqs.~(3.2) -- (3.4) in Ref.~\cite{Gottwald-JCP-2015}.
It becomes apparent at this point that for anharmonic systems this procedure cannot be exploited anymore and one has to use both, the MAF and the MFC, to obtain the spectral density $\hat{\xi}(\omega)$ directly via \Eq{eq:ComputingKernel}. 
This equation, thus, constitutes the generalization of the Fourier method for anharmonic systems.

\subsection{Error Analysis}
\label{sec:Error_analys}
Practical use of \Eq{eq:ComputingKernel} comes along with two numerical issues that have been already encountered in the Fourier method for linear GLEs~\cite{Gottwald-JCP-2015}.
First, the Fourier transforms of the MAF and MFC should be calculated very accurately.
In this respect an integration scheme according to the Simpson 3/8 rule has proven itself reliable. 
Second, the numerical noise in the correlation functions involved should be suppressed for which purpose Gaussian filtering was suggested as the method of choice.
Practically, the correlation functions are multiplied by a Gaussian window function $G(t)=\exp(-t^2/2T)$, where the window width, $T$, might be set to the correlation time of the system
 as a rule of thumb. 
In frequency domain this would lead to a convolution of the signals with a Gaussian of the width $\Delta \omega =1/T$ thereby averaging out the noise.

\begin{figure}[t]
\includegraphics[width=0.99\textwidth]{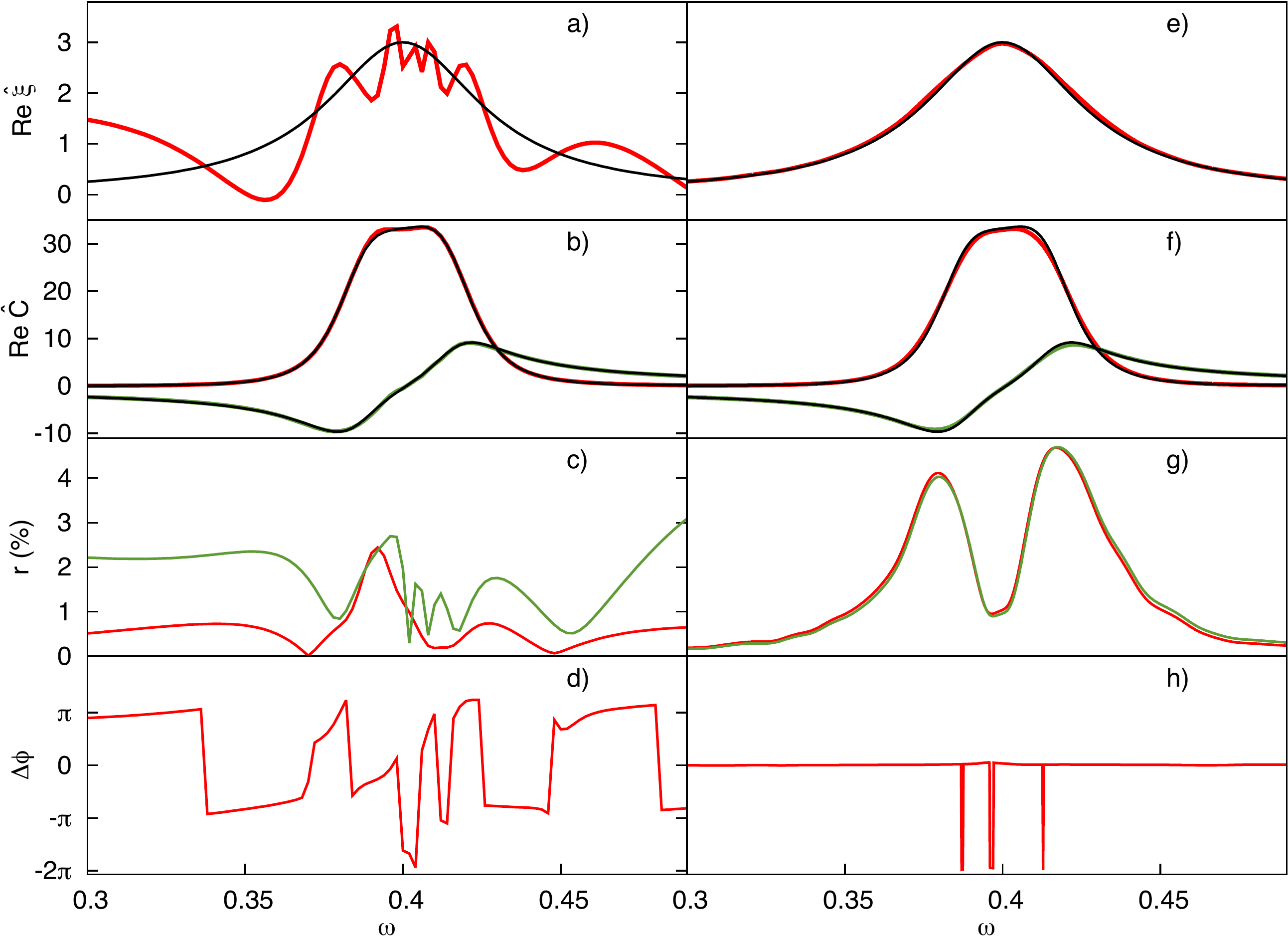}
\caption{
\label{fig:error-spectra}
Numerical results of the fitting procedure (left column) and the Gaussian filtering scheme (right column).
Panels
a,e): Real parts of $\hat{\xi}(\omega)$ with black denoting the true spectral density and red the numerical one,
b, f): real parts of numerical $\hat{C}_{pp}(\omega)$ (red) and $\hat{C}_{pF}(\omega)$ (green) together with exact curves (black), 
c, g):
relative errors $r_{pp}$ (red) and $r_{pF}$ (green),
d, h):
Phase difference $\Delta \phi = \phi_{pF}-\phi_{pp}$. 
}
\end{figure}

In what follows, we demonstrate that the resulting numerical errors are very sensitive to the strategy chosen for tackling them.
In principle, one could opt for a noise reduction scheme alternative to the Gaussian filtering presented above.
As the noise is usually caused by the non-converged tails of the correlation functions, one could fit the MAF and MFC to  damped sine or cosine functions, which suite the natural shape of a typical time correlation function.
This sounds generally appealing as, once the fit is established, no additional numerical errors are introduced and the spectral density can be expressed via \Eq{eq:ComputingKernel} analytically due to the known Fourier transforms of the fit functions, see \Eq{eq:fit function}.
To test the applicability of such an alternative fitting procedure we compare the corresponding memory kernels against the ones obtained from the Gaussian filtering technique in \Fig{fig:error-spectra}.
As an example, a harmonic oscillator with unit mass and $\bar{\omega}=0.4$ in a bath described by a memory kernel $\xi(t)=2a^2\exp[-bt]\cos(ct)$ with $a=0.03$, $b=0.03$ and $c=0.4$ has been used.
The MAF and MFC were calculated from GLE trajectories performed via the Colored Noise thermostat protocol~\cite{Ceriotti2010}.
Note that although the Fourier method developed here is generally designed for anharmonic systems, \Eq{eq:ComputingKernel} remains still valid for harmonic ones.
Choosing a harmonic test system is advantageous because the Fourier-transformed MAFs and MFCs can be computed analytically, see \ref{app:HO_tcfs},
which allows one to calculate numerical errors exactly.
In the fitting procedure the MAF and MFC are fitted to superpositions of
\begin{equation}
\label{eq:fit_f_t}
f(t)=a_1 \exp[-b_1t]\cos(c_1t) + a_2 \exp[-b_2t] \sin(c_2t)
\enspace .
\end{equation}
Surprisingly, the fitting procedure turns out to produce very bad spectral densities,
panel a)
in \Fig{fig:error-spectra}, although the noise reduction in the MAF and MFC is equally successful, see panels b) and f) therein,
and the relative fit errors, panels c) and g), are similarly small as those in the Gaussian filtering scheme.
In contrast, the spectral densities provided by the Gaussian filtering scheme turn out to be very accurate, see panel e).
The reason for these discrepancies becomes evident upon performing a comprehensive error analysis presented in the remainder of this section.

Here we only sketch the derivation for the error of the spectral density, whereas it is detailed in \ref{app:error}.
The error analysis starts with partitioning exact MAF and MFC in \Eq{eq:ComputingKernel} into
\begin{equation}
\hat{C}_{pp/pF}^\mathrm{(exact)}(\omega)=\hat{C}_{pp/pF}^{\mathrm{(num)}}(\omega) + \epsilon_{pp/pF}(\omega)
\enspace ,
\end{equation}
where $\hat{C}_{pp/pF}^{\mathrm{(num)}}(\omega)$ denotes the numerically obtained function and $\epsilon_{pp/pF}(\omega)$ stands for its numerical error that is for the deviation from the exact one.
Then a first-order Taylor expansion with respect to $\epsilon_{pp/pF}(\omega)$ in \Eq{eq:ComputingKernel} is performed
and the relative errors, \Eq{eq:rel_err}, represented in Euler form, 
$r_{pp/pF}(\omega) \e^{i\phi_{pp/pF}(\omega)} \equiv \epsilon_{pp/pF}(\omega) / \hat{C}^{\mathrm{(num)}}_{pp/pF}(\omega)$ are introduced.
After performing some algebra, the error of the memory kernel can be written down as
\begin{equation}
\label{eq:KernelError}
\epsilon_{\xi}(\omega)  \equiv  \left | \hat \xi ^{\mathrm{(exact)}} (\omega)-\hat \xi^{\mathrm{(num)}}(\omega) \right| 
 =  \left | \frac{\hat{C}^{\mathrm{(num)}}_{pF}(\omega)}{\hat{C}^{\mathrm{(num)}}_{pp}(\omega)} \right |
    \left | r_{pF}(\omega) - r_{pp}(\omega)\e^{i\Delta \phi(\omega)} \right |
\enspace,
\end{equation}
where $\Delta \phi(\omega) = \phi_{pp}(\omega)-\phi_{pF}(\omega)$ is the phase difference of the relative errors.

Equation (\ref{eq:KernelError}) shows that the error in the spectral density
can cancel or accumulate depending on the phase difference $\Delta \phi (\omega)$.
Let us assume that the error magnitudes, $r_{pp}(\omega)$ and $r_{pF}(\omega)$ are similar.
Then the error cancellation is supported if the phase difference comes close to a multiple of $2\pi$ whereas the error accumulates if the phase difference is close to an odd multiple of $\pi$.
One can see from \Fig{fig:error-spectra} that the error magnitudes for Gaussian filtering are almost identical (panel g), whereas the phase differences are strictly $0$ or $-2\pi$ (panel h) thereby leading to an almost perfect error cancellation.
In contrast, the phases provided by the fit procedure, panel d), 
often lie close to $\pm\pi$
thereby leading to strong error accumulation.
Since this is not the case everywhere in the resonant region, one can conclude that there is no indication of a systematic error cancellation, in fact, the error rather accumulates.
It is worth noting that since the Fourier method for \textit{linear} GLEs operates exclusively with the MAF, there is no error accumulation from the outset.
Hence, the fit procedure would be equally applicable for this particular case.

All in all, it turns out that controlling the phases is important for a frequency domain procedure to yield accurate spectral densities. 
The Gaussian filtering technique constitutes the method of choice as it seems to offer a phase control that supports error cancellation.
Although it is not possible to perform such an analysis for an arbitrary system due to the absence of exact solutions, the general success of this Gaussian filtering technique for other systems is manifested below, see \Sec{sec:applications}, as well as in our  recent publication, where this method has been applied to solute dynamics in liquid solvent environments~\cite{Gottwald-JPCL-2015}.

The developed method can be summed up by the following steps
\begin{enumerate}
\item{Calculate the MAF and MFC (using the bare system force $F(x)=-\partial_x \VS(x)$) from explicit MD simulations with sufficient convergence (system-dependent)}
\item{Transform the MAF and MFC into frequency domain using a sufficiently accurate integrator, e.g.\ Simpson 3/8 rule, and performing Gaussian filtering, see \Sec{sec:Error_analys}.
%
For the Gaussian window width take the correlation time (estimated from the MAF) as a starting guess}
\item{Calculate the spectral density $\hat{\xi}(\omega)$ according to \Eq{eq:ComputingKernel}}
\item{If the CL model is employed, the renormalization term is accessible from the real part of $\hat{\xi}(\omega)$ according to \Eq{eq:renormalized Potential}}
\end{enumerate}

\section{Simulation Details}
\label{sec:comp_det}

The proposed Fourier method has been applied to the (anharmonic) vibrational dynamics of an I$_2$ molecule in an atomic argon environment. 
The I$-$I interaction is described by a Morse potential whereas the Ar$-$I and Ar$-$Ar interactions are given by Lennard-Jones potentials.
The corresponding parameters, developed for the electronic ground state, have been adopted from Ref.~\cite{Potter1992}.
The system has been investigated at a temperature of $300\,$K in two highly compressed states corresponding to a liquid and solid regime at densities 1592.7 and $2888.6\,$kg$\cdot$m$^{-3}$.
These rather unusual conditions have been chosen for two basic reasons.
On one hand, the investigated vibrational dynamics should be sufficiently anharmonic in order to demonstrate the success of our parametrization method constructed for anharmonic dynamics.
For I$_2$ this implies a sufficiently high temperature, here set to $300\,$K.
On the other hand, we expect the CL model to be a successful description if the real environment is sufficiently harmonic,
which is likely the case in the solid regime.
At a temperature of $300\,$K this amounts to a very high compression.
Note that the 
reliability of the parameters in Ref.~\cite{Potter1992}, that have been optimized for ambient conditions, might be  questionable at such a high compression. 
Nonetheless, the aim of this paper is to demonstrate the success of the proposed method rather than
to investigate the physical properties of a real system at extreme conditions.
Thus,  for the present purpose,  we have employed these parameters without questioning their applicability.

The investigated system has been comprised of a single I$_2$ molecule and 862 Ar atoms in a cubic periodic box of $3.3\,$nm length for the liquid phase and of $2.706\,$nm length for the solid phase.
As an initial configuration, an fcc-lattice of Ar atoms has been prepared.
Two Ar atoms have been replaced by the I$_2$ molecule with the
equilibrium bond length thereby producing a lattice defect.
A $100\,$ps equilibration run with the 
time step of $1\,$fs has been performed to prepare the entire system at a temperature of $300\,$K. 
All explicit MD simulations have been executed
with the GROMACS program package (Version~5.0.1 (double precision))~\cite{GROMACS}.
For calculating vibrational spectra, a set of
$NVE$ trajectories, each $6\,$ps long (time step $1\,$fs), has been started from
uncorrelated initial conditions sampled from an $NVT$ ensemble. 
For controlling the temperature in the equilibration and $NVT$ runs we have used a Langevin thermostat.
MAFs and MFCs
conjugate to the I$-$I bond length have been Fourier-transformed to yield the spectra
according to the procedure described in \Sec{sec:Fourier}. 
In order to achieve convergence, 1000 trajectories have been employed.
Note that although the I$_2$ molecule is IR-inactive the spectra calculated this way can still be interpreted as Raman spectra, if the polarizability is approximated as being proportional to the bond length.

For GLE simulations we have adopted the method of Colored Noise thermostats~\cite{Ceriotti2009a,Ceriotti2010,Morrone2011}.
Since this method can only deal with memory kernels given as a superposition of functions
\begin{equation}
g(t)\equiv2a^2 e^{-bt}\cos(ct) \enspace,
\label{eq: colored noise kernel}
\end{equation}
the spectral densities have been fitted to superpositions of 
\begin{equation}
\label{eq:fit function}
\Re \hat{g}(\omega)=a^2b\cdot \left [ \frac{1}{b^2 + (c-\omega)^2} + \frac{1}{b^2 + (c+\omega)^2} \right ]
\enspace ,
\end{equation}
being the half-sided Fourier transforms of that in \Eq{eq: colored noise kernel}. Usually, 10-20 fit functions are necessary to obtain a good fit.
All numerical parameters for the time step, length, number of trajectories,
etc.\ have been the same as in the explicit MD simulations.

On top, we have performed the same checks of the model assumptions as in our previous publications~\cite{Gottwald-JPCL-2015,Gottwald-JCP-2015}.
The assumptions to be checked are the linearity of the system-bath coupling on the system side, see \Eq{eq:CL Hamiltonian}, the Gaussian statistics of the noise, and the independence of the computed spectral density on the system potential $\VS(x)$.
The linearity of the coupling on the system side can be verified via explicit MD simulations in a straightforward manner.
Here, we sampled $5000$
uncorrelated configurations and varied the system coordinate, $x$,
 within the range accessible due to its thermal fluctuations
keeping all bath coordinates fixed.
The system-bath coupling $V_{\mathrm{S-B}}(x)$ probed this way was least-squares fitted to linear functions $\phi(x)=ax+b$, and quadratic functions $\phi(x)=ax^2+bx+c$ for comparison.
In order to quantify the fit error $r$ with respect to a reasonable scale, we considered the relative deviation 
\begin{equation}
\label{eq:lin-fit-dev}
r(x)=\frac{\phi(x)-V_{\mathrm{S-B}}(x)}{ |V_{\mathrm{S-B}}(x_i)-V_{\mathrm{S-B}}(x_f)|}
\enspace ,
\end{equation}
with $x_i^{}/x_f^{}$ being the initial/final values of the probed range, respectively.
The 
Gaussianity
of the noise, $R(t)$, can be checked by fixing the system's bond length, $x$, to the equilibrium distance of the potential and calculating the distribution function, $f(R)$, of the environmental forces acting on the bond.
For a homonuclear diatomic, the noise is calculated as
\begin{equation}
R=\frac{1}{2} \vec{n} \left [ \vec{F}_1 -\vec{F}_2 \right ] \enspace ,
\end{equation}
where $\vec{F}_{1/2}$ are the forces acting on the individual iodine atoms and $\vec{n}$ is the bond vector.
Here, 500 trajectories of 6ps length have been used to bin the noise. The I$-$I bond length has been fixed using the LINCS algorithm implemented in GROMACS.
The independence of the spectral density on the system potential is checked by substituting the Morse potential by its harmonic approximation and comparing the resulting spectral densities.
For the harmonic system dynamics the same numerical setup has been employed as for the anharmonic case.

\section{Results \label{sec:applications}}

As it was discussed in the Introduction, it is desirable to find systems that are fairly similar to the CL model, 
such that they could be mapped onto it.
One such example is the A$_2$ in A model~\cite{Berne1990,Gottwald-JCP-2015}, which can hardly be connected to a realistic setup.
The intrinsically harmonic structure of the MBO/CL model suggests that its realistic analogues can be found among solids or on surfaces.
Thus, we here employ an I$_2$ in atomic argon in the solid regime.
The liquid regime has been also studied for comparison.
To check the applicability of the mapping onto the CL/MBO models we have compared the vibrational spectra, i.e.\ the Fourier transform of the MAFs, obtained from explicit MD simulations with the ones resulting from the GLE simulations.
For the latter, the spectral densities computed from the MD data according to the scheme proposed in \Sec{sec:method} have been used.
\begin{figure}[t]
\includegraphics[width=0.99\columnwidth]{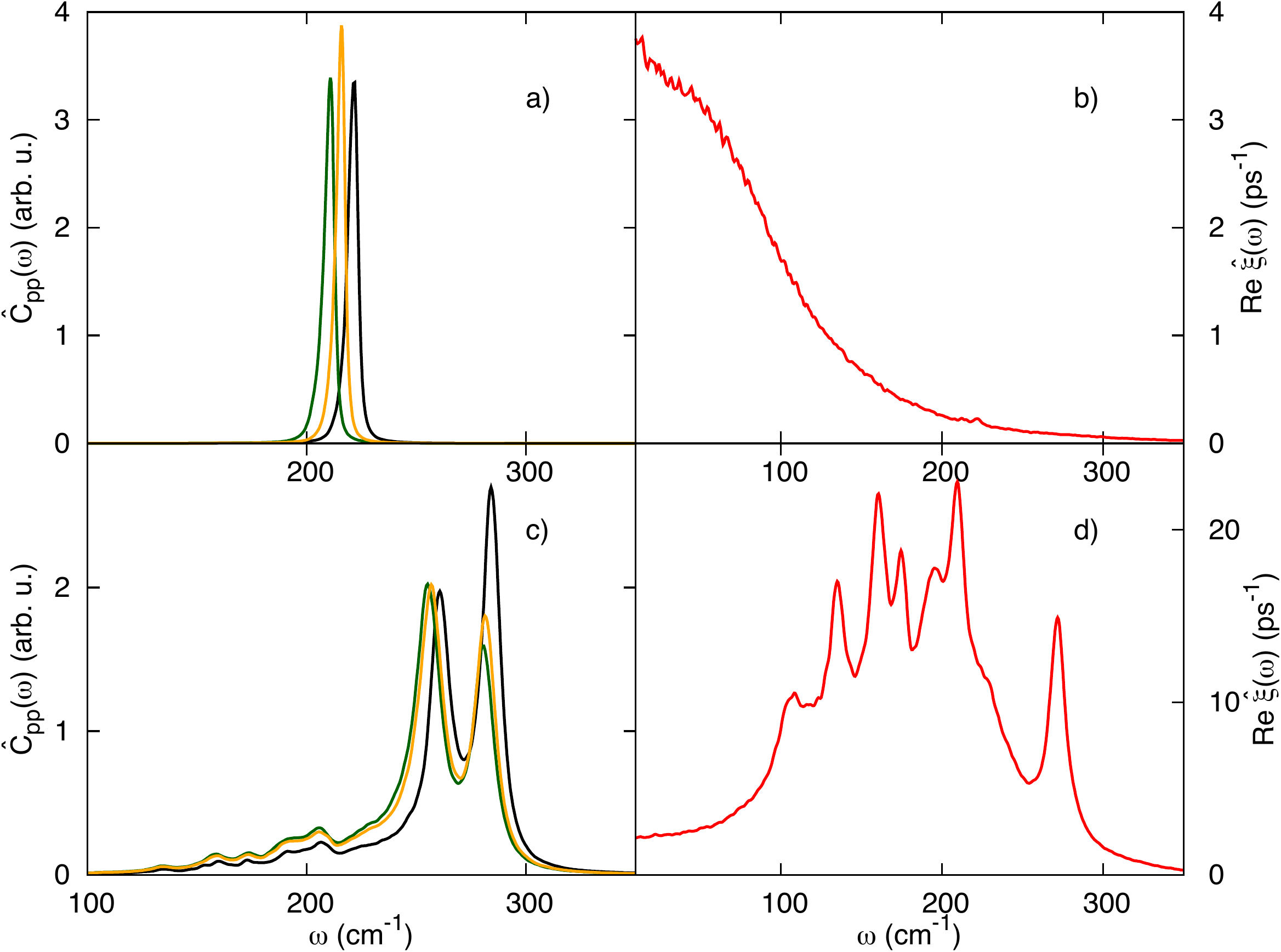}
\caption{
\label{fig:results}
Spectra (left column) and spectral densities (right column) for the liquid (upper row) and for the solid (lower row) I$_2$ in argon at 300K. The spectra are given in black (explicit MD result), yellow (MBO model) and green (CL model).
}
\end{figure}
To reiterate, we are not aiming at reproducing the physical properties of an iodine molecule, but rather at reproducing the explicit MD results for a model system.

In \Fig{fig:results} the spectra (left column) and the corresponding spectral densities (right column) of the investigated system are shown  for the liquid (upper row) and the solid (lower row) regimes.
Starting the discussion with the liquid case, we obtain an unstructured, slightly asymmetric spectral line that is centered at $221\,$\cm\ from the explicit MD simulations (black curve, panel a).
The full width at half maximum (FWHM) is  $5.6\,$\cm; thus, the line broadening is rather small.
Looking at the spectral density (panel b) one observes a broad, 
Debye-type profile
with a maximum at zero-frequency.
In the frequency region that is resonant with the system frequency, the coupling is rather small which implies a large dephasing time and hence explains the small spectral width observed.
This qualitative behavior is reproduced via the corresponding GLE simulations for the CL (green curve) and the MBO (yellow curve) models.
However, one observes that the explicit MD result is blue-shifted for about $10\,$\cm\ compared to the CL result and $6\,$\cm\ compared to the MBO result.
We assign these blue-shifts to the high pressure that is present in the explicit MD simulations. 
This pressure compresses the I$-$I interaction potential thereby yielding a higher frequency compared to the unperturbed case.
This net compression is not accounted for in the CL/MBO models as the average of the noise term is zero by construction.
However, it seems to have only minor influence on the spectra as these shifts can be rated as rather small.
In order to compare to other systems,
the  GLE spectra of the ionic liquid investigated in Ref.~\cite{Gottwald-JPCL-2015} were red-shifted by about $10\,$\cm\ (CL) and blue-shifted by $32\,$\cm\ (MBO) from the explicit MD results.
For the aqueous systems studied therein, the CL/MBO model description failed most significantly and the shifts were $97\,$\cm\ (CL) and $254\,$\cm\ (MBO) to the blue.

In the solid regime, the MD spectrum is very broad and shows two characteristic peaks at 284 and $260\,$\cm\ (black curve, panel c). 
The corresponding spectral density (panel d) is covering a frequency range from $80 \,$\cm\ to $300\,$\cm\ and constitutes a peaked pattern that can be assigned to lattice vibrations of the Ar crystal.
The peak around $280\,$\cm\ implies a strong resonant coupling of the I$-$I vibration to the phonon bath which causes the characteristic peak splitting in the spectrum.
These features are well-reproduced in the GLE simulations corresponding to the CL (green curve) and MBO (yellow curve) models.
As in the liquid regime, a blue-shift of the same order of magnitude of the explicit MD spectra is observed and can, again, be explained by the high pressure.
Interestingly, the deviations between the MBO and CL models themselves are negligible in the case of a solid.
Overall, the GLE simulations according to the CL and MBO models lead to satisfactory vibrational spectra. 
Especially in the solid regime, where the spectrum shows a more complicated structure, the MBO/CL models cover the main effects with good accuracy.
Although the MBO results are closer to the explicit MD results than the CL ones, especially in the liquid regime, we explicitly have to warn the reader about concluding that the MBO model is generally to be preferred to the CL model.
This conclusion would not be correct as there exist systems where the CL results are closer to the MD ones, see~\cite{Gottwald-JPCL-2015}.
In general, the renormalization term in the CL model shifts the system frequency to the red, thereby making a description of systems,
where a bath causes a blue shift with respect to gas phase, worse.
Thus, we doubt that a rule on whether to choose the MBO or CL model, i.e.\ to keep or not to keep the frequency renormalization term in the GLE, can be given in general.
\begin{figure}[t]
\includegraphics[width=0.99\columnwidth]{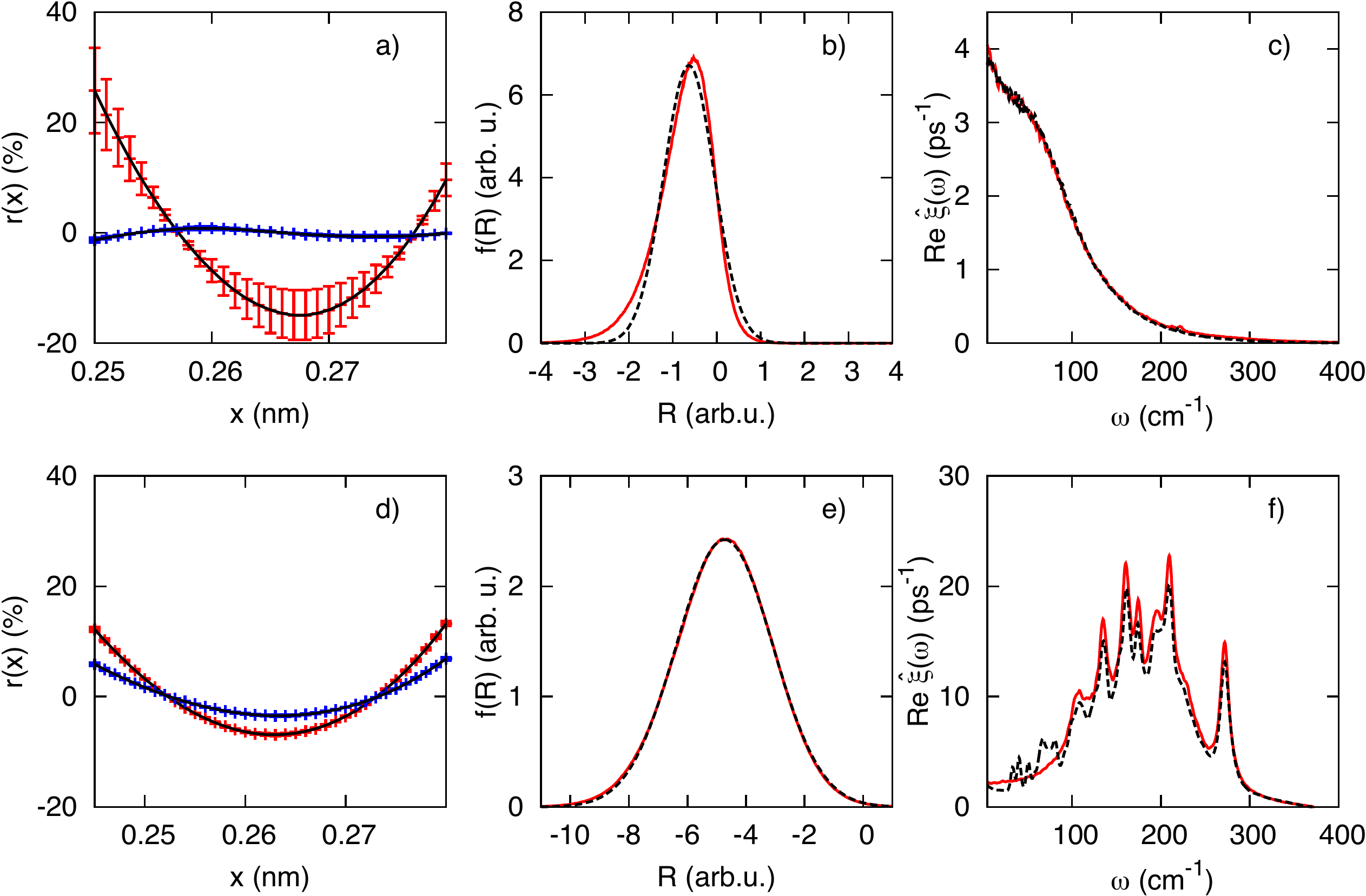}
\caption{
\label{fig:CL-check}
The checks of the CL/MBO model assumptions for I$_2$ in Argon at $300\,$K is shown for the liquid regime (upper row) and the solid regime (lower row).
The fit errors (left panels) are displayed in error bars for the linear (red) and quadratic (blue) fits.
The noise distribution functions (middle panels) from MD simulations (red curves) are compared against a fit to Gaussian functions (black dashed curves) $f(R)=A \exp(-B\cdot (R-C)^2)$. 
The spectral densities (right panels) obtained from an anharmonic system potential (red curves) are plotted together with the ones obtained for a harmonic potential (black dashed curves). 
}
\end{figure}

One may ask, if it is possible to formulate criteria stating whether the system would or would not be mapped onto the
CL/MBO model successfully. 
Although we think that performing the direct comparison of GLE and MD results on the basis of observables, like vibrational spectra, is the most straightforward way to check the applicability, it might still be desirable to have such \textit{a priori} criteria which can be applied based exclusively on the MD data.
This is particularly convenient, when the machinery to perform GLE simulations is not available or when no further investigations on the basis of classical GLE dynamics are intended.
Possible candidates for such criteria are the linearity of the system-bath coupling on the system side, the Gaussianity of the random force, $R(t)$, and the independence of the spectral density on the system potential~\cite{Gottwald-JPCL-2015, Gottwald-JCP-2015}.
Our previous analysis yielded that the aqueous systems considered in \cite{Gottwald-JPCL-2015} exhibited almost linear coupling on the system side (the deviations were smaller that 2\%), whereas for the ionic liquid considered therein it was completely non-linear (the deviations were up to $\approx 40\%$).
The Gaussianity of the noise revealed the opposite trend, that is the noise was not centered at zero and fairly asymmetric for aqueous systems, whereas it was almost Gaussian and zero-centered for the ionic liquid considered.
Since the GLE spectrum for the ionic liquid was closer to the MD results compared to those of the aqueous systems, which were miserably wrong in shape and peak position,
we concluded that Gaussianity might be more important than the linearity of the coupling on the system side~\cite{Gottwald-JPCL-2015}.
This trend was fully confirmed by considering the A$_2$ in A model system in Ref.~\cite{Gottwald-JCP-2015}.
Additionally, the independence of the spectral density on the system potential turned out to be a reliable criterion for the applicability of the mapping onto the CL/MBO models. 
Whereas the spectral density for A$_2$ in A  was nearly potential-independent a significant dependence on the system potential has been observed for the aqueous systems~\cite{Gottwald-JCP-2015}, which is an indication that these are not of the CL form correlating nicely to the bad quality of the corresponding GLE spectra.

The same analysis is presented in \Fig{fig:CL-check} for the I$_2$ in argon investigated here.
Starting the discussion with the linearity of the coupling on the system side (panels a) and d) therein), one notices rather large fit errors of about 25\% in the liquid regime and about 10\% in the solid regime.
Performing a quadratic fit shows an improvement in both cases, which underlines that the linearity assumption of the coupling does not hold for the investigated system.
In general, having a higher non-linearity in the liquid regime is expected as the motions are of larger amplitude than in the solid regime.

One might wonder, whether a quadratic dependence of the system-bath coupling on the bond length indicates that the MBO model is to be preferred to the CL model, as, in the MBO model, the quadratic counter-term is viewed as a part of the system-bath coupling, see \Eq{eq:MBO Hamiltonian}.
Then, this quadratic term should coincide with the renormalization term, \Eq{eq:renormalized Potential}.
However, the renormalization term could not be  responsible for this quadratic behaviour as it turned out to be orders of magnitude different.
Furthermore, the renormalization term does not depend on the particular bath configuration, whereas the quadratic fit coefficient exhibits strong dependence on bath coordinates.

The assumption of a Gaussian distributed noise (panels b and e) is sufficiently verified for the solid regime.
In the liquid regime one observes a slightly asymmetric profile that still fits well a Gaussian shape.
%
In both cases the mean of the noise is not centered at zero but at negative values.
This can again be attributed to the high pressure which, on average, produces environmental forces that shrink the I$-$I bond.
This, however, results only in small shifts in the spectra, see \Fig{fig:results}, 
implying that the relation between the induced pressure and the spectral shifts is far from trivial.
Finally, the spectral densities (panels c and f) obtained for the harmonic (black dashed curves) and for the anharmonic (red curves) potentials do not deviate remarkably in both the liquid and solid regimes.

It can be stated that the trends observed in our earlier publications are confirmed here as well. 
In particular, the success of the MBO/CL model to mimic vibrational spectra does not seem to be sensitive to the linearity of the system-bath coupling on the system side. 
In turn, the Gaussian shape of the noise distribution as well as the independence of the spectral density on the system potential seem to offer reliable criteria to predict the applicability of the mapping onto the MBO/CL models.

%

\section{Conclusion \label{sec:conclusion}}

%
In this paper, we have extended the Fourier-based scheme suggested in~\cite{Gottwald-JCP-2015}
to parameterizing a non-linear GLE according to the CL/MBO model with anharmonic system potentials from explicit MD simulations.
The accuracy of the resulting procedure turned out to be very sensitive to numerical errors of the input data.
Based on the detailed error analysis we have found that a Gaussian filtering scheme is preferable for a numerical noise reduction as it could limit the error accumulation in an efficient way.
Further, we have shown that, despite the possible inapplicability of the models to the spectroscopy of arbitrary systems in solution due to the invertibility problem~\cite{Gottwald-JPCL-2015},
the mapping of a real system onto the CL and/or the MBO model with \textit{anharmonic} system potentials is possible for a potentially broad class of systems.
Importantly, the successful example shown in this paper demonstrates that the invertibility problem can have exceptional cases where it does not surface and, hence, it does not disqualify using the CL/MBO model as such.
The models should be rather used with caution and a critical assessment of their applicability should be performed before trusting in the results.
To our understanding the ultimate way is to check the occurrence of the invertibility by comparing GLE and MD data. This would also clarify whether the renormalization term should be incorporated or not.
However, the assessment of the applicability can also be based on \textit{a priori} criteria that have been identified to be the Gaussianity of the noise distribution function and the independence of the spectral density on the system potential employed.
These empirical criteria may serve as a handy tool to judge on the applicability, employing exclusively the explicit MD simulation data.

\section*{Acknowledgements}

We would like to thank Sven Karsten for fruitful discussions.
The authors gratefully acknowledge financial support by the Deutsche
Forschungsgemeinschaft (Sfb~652 (O.K.) and IV~171/2-1 (S.D.I.)).

\appendix
\setcounter{section}{0}

\section{Correlation functions for a harmonic oscillator}
\label{app:HO_tcfs}
For a harmonic oscillator with a frequency $\omega_0$, the MAF, $C_{pp}(t)$, and the MFC, $C_{pF}(t)$, (with $F=-m\omega_0^2x$) can be analytically expressed in terms of the spectral density $\hat{\xi}(\omega)$. 
Starting from the Volterra-type integro-differential equation connecting the MAF and MFC, \Eq{eq:Volterraequation-CL},
that reads
\begin{equation}
\dot{C}_{pp}(t)=-m\omega_0^2C_{xp}(t) - \intop_0^{\infty} \xi(t-\tau)C_{pp}(\tau) \diff \tau
\end{equation}
and applying the half-sided Fourier transform for both sides yields
\begin{equation}
i\omega \hat{C}_{pp}(\omega)-C_{pp}(t=0)=-m\omega_0^2\hat{C}_{px}(\omega)-\hat{\xi}(\omega)\hat{C}_{pp}(\omega) \enspace.
\end{equation}
Using the relation
\begin{equation}
\dot{C}_{px}(t)=\frac{C_{pp}(t)}{m} \enspace ,
\end{equation}
which in Fourier space translates into
\begin{equation}
\label{eq:Cpx_Cpp}
i\omega\hat{C}_{px}(\omega)-C_{px}(t=0)=\frac{\hat{C}_{pp}(\omega)}{m} \enspace ,
\end{equation}
results in
\begin{equation}
i\omega \hat{C}_{pp}(\omega)-1=i\frac{\omega_0^2}{\omega}\hat{C}_{pp}(\omega)-\hat{\xi}(\omega)\hat{C}_{pp}(\omega) \enspace,
\end{equation}
where it has been used that $C_{px}(0)=\langle p x \rangle =0$ for the canonical ensemble and it has been assumed that
the MAF is normalized to unity, that is $C_{pp}(t=0)=1$.
Finally one obtains for the MAF
\begin{equation}
\hat{C}_{pp}(\omega)=\frac{\omega}{\omega \hat{\xi}(\omega) +i(\omega^2-\omega_0^2)}
\enspace .
\end{equation}
The MFC can be obtained from the MAF with the help of \Eq{eq:Cpx_Cpp}, since
$\hat{C}_{pF}(\omega)=-m\omega_0^2\hat{C}_{px}(\omega)$.
The resulting expression for the MFC reads
\begin{equation}
\hat{C}_{pF} (\omega)=\frac{i\omega_0^2}{\omega \hat{\xi}(\omega) +i(\omega^2-\omega_0^2)}
\enspace .
\end{equation}

\section{Detailed derivation of numerical error formula}
\label{app:error}
A numerically computed correlation function in the frequency domain can be related to the exact one as 
%
\begin{equation}
\hat{C}^{\mathrm{(exact)}}(\omega)=\hat{C}^{\mathrm{(num)}}(\omega) + \hat\epsilon(\omega)
\enspace ,
\end{equation}
where $\hat{C}^{\mathrm{(num)}}(\omega)$ denotes the numerically obtained function and $\hat{\epsilon}(\omega)$ the numerical error.
With the help of \Eq{eq:ComputingKernel}, the exact spectral density becomes
\begin{equation}
\hat{\xi} ^{\mathrm{(exact)}}= \frac{1+\hat{C}^{\mathrm{(num)}}_{pF}+\hat \epsilon_{pF}}{\hat{C}^{\mathrm{(num)}}_{pp}+\hat \epsilon_{pp}} -i\omega \enspace .
\end{equation}
%
Defining the relative errors as
\begin{equation}
\label{eq:rel_err}
\hat{r}_{pp/pF}(\omega) \equiv \frac{\hat{\epsilon}_{pp/pF}(\omega)}{\hat{C}^{\mathrm{(num)}}_{pp/pF}(\omega)}
\end{equation}
and performing a first order Taylor expansion with respect to the errors yields
\begin{eqnarray}
\hat{\xi}^{\mathrm{(exact)}}(\omega) & = &
  \frac{1+\hat{C}^{\mathrm{(num)}}_{pF}(\omega)}{\hat{C}^{\mathrm{(num)}}_{pp}(\omega)} -i\omega +
  \frac{\hat{\epsilon}_{pF}(\omega)}{\hat{C}^{\mathrm{(num)}}_{pp}(\omega)} - \frac{1+\hat{C}^{\mathrm{(num)}}_{pF}(\omega)}{[\hat{C}^{\mathrm{(num)}}_{pp}(\omega)]^2} \hat{\epsilon}_{pp}(\omega) \nonumber \\ 
 & = & \hat{\xi}^{\mathrm{(num)}}(\omega) + \frac{\hat{C}^{\mathrm{(num)}}_{pF}(\omega)}{\hat{C}^{\mathrm{(num)}}_{pp}(\omega)} (\hat r_{pF}(\omega) - \hat r_{pp}(\omega)) - \frac{\hat r_{pp}(\omega)}{\hat{C}^{\mathrm{(num)}}_{pp}(\omega)} 
\enspace .
\end{eqnarray}
Assuming that the relative errors are small, one can take $\frac{\hat r_{pp}(\omega)}{\hat{C}^{\mathrm{(num)}}_{pp}(\omega)} \approx 0$ thereby neglecting the last term.
Further, writing the relative errors in Euler form $\hat r_{pp/pF}=r_{pp/pF} \e^{i\phi_{pp/pF}}$, one arrives at 
the error formula used in the main text, \Eq{eq:KernelError}.
\begin{eqnarray}
\hat \epsilon_{\xi} & \equiv & \left | \hat \xi ^{\mathrm{(exact)}}-\hat \xi^{\mathrm{(num)}} \right| \nonumber \\ 
& = & \left | \frac{\hat{C}^{\mathrm{(num)}}_{pF}}{\hat{C}^{\mathrm{(num)}}_{pp}} \right | \left | r_{pF} - r_{pp} \e^{i\Delta \phi} \right | \enspace,
\end{eqnarray}
with the phase difference $\Delta \phi = \phi_{pp}-\phi_{pF}$.

\bibliographystyle{iopart-num}
\bibliography{./GLE}
\end{document}